\documentclass[nologo,11pt,a4paper]{ETHpaper}
\usepackage{verbatim}
\usepackage{graphicx}
\usepackage{subfigure}

\usepackage{graphicx,epsfig}
\usepackage{times}
\usepackage{graphics,dcolumn,bm,fleqn,epic,eepic}
\usepackage{amssymb,amsmath,multirow,rotate,color}

\usepackage{color}

\begin{document}

\title{Diversity-induced resonance in the response to social norms}

\titlealternative{Diversity-induced resonance in the response to social norms}

\author{Claudio J.\ Tessone$^1$, Angel S\'anchez$^{2,3}$, Frank Schweitzer$^1$}

\authoralternative{C.J.\ Tessone, A.~S\'anchez, F.~Schweitzer}

\address{$^1$ Chair for Systems Design, ETH Zurich, Weinbergstrasse 58, CH-8092 Z\"urich, Switzerland \newline
$^2$ Grupo Interdisciplinar de Sistemas Complejos (GISC),
  Departamento de Matem\'aticas, Universidad Carlos III, Legan\'es
  (Madrid) E-28933, Spain \newline
$^3$ Institute for Biocomputation and Physics of Complex
Systems (BIFI), University of Zaragoza, Zaragoza 50009, Spain}

\date{\today}

\reference{Submitted (2012). }

\www{\url{http://www.sg.ethz.ch}}

\makeframing


\maketitle

\begin{center}
 \date{\today}
\end{center}

\begin{abstract}
  In this paper we focus on diversity-induced resonance, which was
  recently found in bistable, excitable and other physical systems. We
  study the appearance of this phenomenon in a purely economic
  model of cooperating and defecting agents. Agent's contribution to a
  public good is seen as a social norm. So defecting agents face a social
  pressure, which decreases if free-riding becomes widespread.
  In this model, diversity among agents naturally appears because of the different sensitivity towards the
  social norm.
  We
  study the evolution of cooperation as a  response to the social norm (i)
  for the replicator dynamics, and (ii) for the logit dynamics by means
  of numerical simulations.
  Diversity-induced resonance is observed as
  a maximum in the response of  agents to changes in the social
  norm as a function of the degree of heterogeneity in the population. We provide an analytical, mean-field approach for the logit dynamics
  and find very good agreement with the simulations.
From a socio-economic perspective, our results show that,
  counter-intuitively, diversity in the individual sensitivity to social norms may result in a
  society that better follows such norms as a whole, even if part of the
  population is less prone to follow them.
\end{abstract}

\section{Introduction}

The ever-increasing interest by physicists to contribute to understanding
collective phenomena in social systems \cite{castellano:2009} has mostly
concentrated around highly stylized models, often directly borrowed from
physics, using vague plausibility arguments to justify their social
context \cite{stauffer:2003}. In this paper, we follow a less common
route, namely to work with a model which is established in, and directly
taken from, the social sciences. It studies the effect of social norms on
the emergence of cooperation. We study its dynamics from the
physical perspective of diversity-induced resonance, to shed new light on
sustainable cooperation in a society where some fractions do not adhere
to support it.

In a system consisting of distinct and non-identical elements,
diversity-induced resonance can be defined as the appearance of a maximum
response to an external signal, dependent on the degree of
diversity. This phenomenon was first reported in Ref.~\cite{tessone:2006}, in the context of coupled bistable or excitable
systems that are subject to a sub-threshold signal. It was shown that
there is an optimum level of the diversity (quenched noise) of the
coupled units that maximizes the
response to the signal. Subsequent works \cite{tessone:2007, toral:2007, Chen2009, Chen2009a,
  Komin2010, Calisto2010, Perez2010, WANG2010, Wu2011, McDonnell2011}
showed that similar behavior can be observed in other physical systems,
thus reinforcing the notion that this type of resonance can be a quite
general phenomenon. In fact, diversity-induced resonance was also shown
to appear in models related to sociophysics: It was found in discrete
models of opinion formation \cite{tessone:2009} --such as the Galam model
\cite{galam:1997} (related to the random-field Ising model at zero
temperature \cite{sethna:1993}-- and in continuous ones
\cite{VazMartins2010}, of which the Deffuant model \cite{Deffuant2000} is
a paradigmatic example. In all cases, the average opinion synchronizes to
external signals or influences when the diversity in the preferred
opinions attains an optimum value.  In a broader context, diversity-induced
resonance can be generalized to other sources of {\em disorder} in the
internal dynamics of the system constituents. Interestingly, even
repulsive and evolving patterns of interactions can trigger a common collective behavior, be it
synchronization \cite{Tessone2008,Tessone2012}, an amplification of
an external signal \cite{VazMartins2010,PhysRevE.81.041103} or a
nonlinear increase in the volatility of the global dynamics
\cite{PhysRevE.85.011150}.  In a social context, these repulsive
interactions would represent contrarians, i.e.  individuals that oppose
any type of consensus
\cite{GALAM2004,Stauffer2004558} or that intend to destabilize the system
itself, such as the joker-like players studied in the context of social
dilemmas \cite{Arenas2011113}.

The research reported here generalizes diversity-induced resonance by
demonstrating its appearance in a purely economic model of social norms
and their effect on cooperation \cite{spichtig:2011}. Instead of relying
on a model rooted in physics, we study an established model from
the economics literature in which diversity and external driving are
introduced based on economic considerations.
In this model, diversity appears naturally as an idiosyncratic propensity to follow a social norm.
We demonstrate for this
model that there is an optimal range of diversity, which leads the society to follow such norms as a
whole. Different from the setup of diversity-induced resonance
models usually studied in the physics literature, in this case diversity appears in a
multiplicative manner and its dynamics are given by approaches typical of evolutionary
game theory.

The paper is organized as follows: Section \ref{sec:2} presents our model
and its economic context. Section \ref{sec:3} summarizes our simulation
results, obtained for two different types of evolutionary dynamics, to
demonstrate the robustness of the observations.  To better understand the
origin of the collective dynamics, we present our findings for three
levels of increasing modeling complexity, without and with diversity, and
with external forcing.  Subsequently, Sec.\ \ref{sec:nueva} improves our
understanding by means of an analytical approach for the stationary level of cooperation, whereas Sec.\ \ref{sec:otra}
investigates the response to the external signal.  Finally, Section
\ref{sec:4} summarizes our conclusions and discusses the implications of
this work.

\section{Model}
\label{sec:2}
\subsection{Economic context}

In this paper, we model {\em conditional cooperation}, a phenomenon observed in
many human interactions. This term was
introduced by Keser and van Winden \cite{keser:2000} and Fischbacher {\em
  et al.}\ \cite{fischbacher:2001} to refer to the fact that people often
condition their cooperation on the cooperativeness of others or on their
beliefs about others' behavior. In the specific context of Prisoner's
Dilemma \cite{rapoport:1966,axelrod:1984} or Public Goods games
\cite{kagel:1995}, this means that people are ready to contribute more to
the common welfare if others contribute as well. Furthermore, this
willingness increases with the number of contributors in the game. There
is a large body of experimental evidence supporting the existence of this
type of behavior \cite{gachter:2007}, even in structured populations
\cite{grujic:2010,grujic:2012,gracia-lazaro:2012}. It is only consistent to ask (i) for a deeper
theoretical understanding of these observations and (ii) their
consequences for economic reasoning.  The first question is partly
answered by the theory of social preferences \cite{fehr:2006}, that
posits that non-monetary contributions to the utility function arise from
social considerations, such as, e.g., inequity aversion or reciprocity.
It has been argued that social preferences arise through social norms,
i.e., rules of conduct that are enforced by internal or external
sanctions \cite{coleman:1990}. Explanations for the emergence and
robustness of such norms in evolutionary terms have been advanced
\cite{galan:2005,mengel:2008, spichtig:2011}, thus closing the rationale
to explain conditional cooperation in terms of social preferences.

In this paper, we focus on the issue of norms and on the consequences of having a
diverse population of conditional cooperators interacting in a Public
Goods setup. Thus, we investigate how diversity influences the response to
exogenous efforts to promote cooperation through social norms.  Following
Spichtig and Traxler \cite{spichtig:2011}, we consider that a norm
against free-riding is enforced (internally or externally). This is
achieved by adding a contribution to the utility function such that
free-riding (i.e., not contributing to the public good while benefiting
from it) is heavily punished when rare, but the punishment weakens as
free-riding becomes more abundant in the population. This norm leads to
conditional cooperation because of more willingness to cooperate when the
population is mostly cooperative, and the propensity to cooperate
decreases if less participants cooperate.

In the above context, we address the following question: How does the
behavior of the population change if the social norm responsible for
establishing a conditionally cooperative strategy varies in time?  This
question is important for two reasons. First, social norms are known to
change in time, endogenously or exogenously, in periodic or random
manners \cite{ostrom:2000,posner:2000}. Therefore, it is most important
to understand how those changes affect the observed behavior in order to
assess the stability of cooperative environments. Second, understanding
the response of the population to changes in the current social norm can
help policy makers to design incentives or new norms that lead to more
cooperative outcomes. However, it should also be realized that the effort
of steering the norms towards a preferred direction is costly and, at
some point, the benefit of improving the behavior of the populations may
be lower than that of continuing changing the norm. Therefore, assessing
the optimum amount of effort invested in modifying a given norm is a very
relevant issue. Finally, we will come to the issue of diversity-induced
resonance by considering that the sensitivity to the social norm depends
on the individual through a specific coefficient to be introduced in the
utility function.
In the following, we will show that these issues can be addressed --and are related to-- the phenomenon of diversity induced resonance in this system

\subsection{Model definition}

Let us now implement the ideas above in a well-defined model built on the
original proposal by Spichtig and Traxler \cite{spichtig:2011}.  We
consider a population of $N$ agents which can take one of two
possible (opposite) actions $\sigma_i \in \left\{0, 1 \right\}$, for $i =1\cdots N$. We
assume that ``cooperative'' agents take action $\sigma=1$, this way
contributing to a public good, while ``free riders'' take action
$\sigma=0$ and do not contribute to the public good. Defining the density
of cooperators as $n_c\equiv N_c/N$, and the density of free-riders as
$n_f\equiv 1-n_c = N_f/N$ respectively, the utility (or payoff) function
per agent is defined as
\begin{eqnarray}
\label{payoff-function}
u_i(\sigma_i, \theta_i;n_c)\equiv &  - & c \, \sigma_i
+ \frac{r}{N}\sum_{j=1}^N \sigma_j \nonumber \\ &+ &
  (\sigma_i-1)\, \theta_i s(n_c)
\end{eqnarray}
The first term in Eq.~(\ref{payoff-function}) represents the cost $c$
per agent for providing the public good, which applies only if agent $i$
is cooperative, $\sigma_{i}=1$.  The second term represents the benefit
$r/N$ per agent resulting from the public good. It applies regardless of
the agent's action $\sigma_{i}$. Both terms describe the utility function
of a classical public good game. The third term, new to the model,
describes an additional effect resulting from the existence of a social
norm, or social pressure, to cooperate. Free-riders with $\sigma_{i}=0$
face an (internal or external) sanction \cite{coleman:1990}, which does
not apply for cooperators with $\sigma_{i}=1$.  We assume that the
strength of the social pressure $s(n_c)$ depends on the density of
cooperators. If $n_{c}$ is small, i.e. if free-riding is widespread, then
agents deviating from cooperation may face weaker sanctions. Hence,
$s(n_c)$ is assumed to increase monotonously with $n_c$, with $s(1)>0$
and $\lim_{n_c\to 0}s(n_c)=0$.  In the following, we simply choose a
linear function $s(n_c)=\alpha \, n_c $, with $\alpha> 0$.

Eventually, we consider that not all agents may be prone to social
pressure in the same manner. To cope with this individual sensitivity to
the social norm, we introduce a new variable $\theta$ with realizations
$\theta_i$ drawn from a probability distribution function $g(\theta)$
with mean $\Theta$ and standard deviation $\Delta\theta$.  Note that
negative values of $\theta_{i}$ imply a positive contribution to the {\em
  perceived} agent's utility by violating the social norm. This reflects
the presence of contrarians/jokers \cite{Arenas2011113} in the population
that are willing to go against the system in order to benefit. Such
individuals would more likely not contribute to the public good in
presence of social pressure but, as we will see below, their presence
turns out not to be an obstacle for the general population to conform to
the social norm.

With this utility function, the rational choice of an agent on what action
to take depends on the density of free riders, $n_{c}$ and on her
individual sensitivity, $\theta_{i}$. Introducing $\tilde{c}\equiv
c-r/N$, it is easy to see that agents' decisions can be classified in
three types: (i) agents will always cooperate, $\sigma_i=1$, if
$\theta_i>\tilde{c}/s(N^{-1})$, (ii) agents will always free-ride,
$\sigma_i=0$, if $\theta_i<\tilde{c}/s(1)$, and (iii) agents are
conditional cooperators dependent on the density of free-riders in the
population, i.e. they cooperate if $\tilde{c}/s(N^{-1})> \theta_i>
\tilde{c}/s(1)$.  Note that because of $\lim_{n_c\to 0}s(n_c)=0$, the
criterion for the existence of cooperators is quite tight and often they
will be absent from the population.  Hence, the diversity in the
individual sensitivity $\theta$, precisely the standard deviation $\Delta
\theta$, will play an important role in deciding about the size of the
three groups defined above.  The final level of cooperation (as well as
the influence of the social norm) will to a large extent be governed by
the conditional cooperators.

Finally, we
will consider that the cooperation-fostering norm changes in time, which
is modeled by assuming a time dependence of $\alpha\to\alpha(t)$. This
corresponds to a change of the slope of the social pressure function,
representing periods in history where free-riding is less tolerated than
in others, but it is always tolerated if widespread. If we further assume
that agents can change their action depending on their expected utility,
i.e.~the density of cooperators $n_{c}$ has a dynamics defined like in
the following section, the third term in Eq.~(\ref{payoff-function})
representing the social pressure becomes $(\sigma_i-1) \theta_i
\alpha(t) n_c(t)$. Hence, we have a signal $\alpha(t)$ that changes
over time because of external influences.
In the present paper, for the sake of simplicity, and without altering the main results \cite{PhysRevE.85.011150}, we will consider a periodic change in the amplitude of the social norms.
In absence of
cooperators, the effect of this signal vanishes as well. The diversity in
responding to the signal is given by the individual variables
$(\sigma_i-1)\theta_{i}$, i.e. only free-riders will face the social
pressure, but they are prone to it in a heterogeneous manner.

Studying the model in the setting of diversity-induced resonance allows
us to use standard techniques for quantifying the response of the
population to (for example) a change in the social pressure induced by a
policy change. If the period of the signal $\alpha(t)$ is long enough,
the results of a periodic forcing become equivalent to a one-time
modification. Moreover,  in contrast with previous studies of this phenomenon, the signal enters multiplicatively on the heterogeneous term.

\subsection{Evolutionary dynamics}\label{evol_dyn}

As mentioned above, we implement a dynamics that allow agents to change
their actions dependent on the utility expected. For this dynamics, we
use a standard evolutionary game-theoretical setup with one-shot games,
i.e.  agents have no memory about their previous action.  We consider a
well-mixed population, i.e.  all agents interact together. This is
dynamically equivalent to considering a mean-field version of the public
goods game, already reflected in the sum term in Eq.\
(\ref{payoff-function}). After each round of the game, agents collect
their payoff and subsequently update their strategies according to two
different dynamical rules, which we explain in detail below. From the
various propositions for update rules in the literature
\cite{szabo:2007,roca:2009a}, we have chosen (i) the replicator dynamics
\cite{helbing:1992,schlag:1998}, which is widely used and has a well
defined limit for $N\to\infty$, namely the celebrated replicator equation
\cite{hofbauer:1998,gintis:2009}; and (ii) the logit dynamics
\cite{alos-ferrer:2010}, which allows for the possibility of errors or
mistakes in choosing actions, and whose deterministic limit coincides
with the best-response rule, widely used in economics
\cite{ellison:1993}.

From a socio-economic context, both dynamic rules have a different
interpretation. On the one hand, the replicator dynamics involves some
degree of social interaction (the process is driven by imitation of
successful strategies). On the other hand, the logit dynamics is simply
based on strategic behavior.  By choosing these quite different kinds of
dynamics, we demonstrate the generality and the robustness of the results
presented in this paper.

Regarding the formal description, it is important to notice that the
diversity for $\theta$ introduced in our model no longer allows us to
write down the macroscopic dynamics in terms of a single master equation.
Instead, the system dynamics has to be split into the dynamics of groups
of agents with the same value of $\theta$.  Let $n(\theta,\sigma)
\delta\theta$ be the number of agents with an individual sensitivity in
the interval $[\theta - \delta \theta/2,\theta + \delta \theta/2]$
choosing action $\sigma$ at time $t$ (for simplicity, we also say agents
are in state $\sigma$ at time $t$, i.e. ``state'' refers to
``action''). Then, the rate equation for the density of cooperators with
a sensitivity $\theta$ is given by
\begin{equation}
  \dot n( \theta, 1) = n( \theta, 0)\,  \omega_+ (\theta)
  - n(\theta, 1) \, \omega_- (\theta)
\end{equation}
The transition rates $\omega_+ (\theta)$ ($\omega_- (\theta)$) specify
the overall transition into the state $\sigma=1$
(respectively,~$\sigma=0$) for the two subpopulations with a given
sensitivity $\theta$, but different states.  These transition rates
depend on the dynamic rules chosen and are specified in the following.

\subsubsection{Replicator dynamics}

With this update rule, after every time step all agents revise their
action simultaneously by selecting one neighbor at random, e.g. agent
$j$, and comparing their own payoff $u_{i}$ with their neighbor's payoff,
$u_j$.  If $u_i>u_j$ agent $i$ keeps her action, whereas in the opposite
case it adopts the action of the more successful agent $j$ with a
probability proportional to $(u_j-u_i)$.  Replicator dynamics is purely
imitative, meaning that  actions not present currently in the system  can not appear spontaneously.
This in turn implies that states in which
\emph{all} agents defect or \emph{all} contribute are absorbing
states. In order to let the system leave those absorbing states, we have
introduced noise to the dynamics: with a small probability $\epsilon$ an
agent can switch her action spontaneously at every time step.
Subsequently, all payoffs are reset to zero and a new round of the game
proceeds.

For agents with an individual sensitivity $\theta$, the overall transition
rate towards the opposite state depends on the possible pairings with
agents in the opposite state and equipped with individual
sensitivity $\theta'$. This yields
\begin{eqnarray}
\omega_- (\theta) &=&
\epsilon +  \int \omega_-(\theta | \theta')
\, n(\theta', 0) \, g(\theta')\, d\theta', \label{trans_totm} \\
\omega_+ (\theta) &=&
\epsilon +  \int \omega_+(\theta | \theta')
\, n(\theta', +1) \, g(\theta')\, d\theta', \label{trans_totp}
\end{eqnarray}
where $g(\theta)$ is the distribution function of $\theta$.  The
conditional transition rates $\omega_{+}(\theta | \theta')$,
$\omega_{-}(\theta | \theta')$ are equal to the differences in payoff, if
the payoff of the agent with $\theta'$ is larger, i.e.
\begin{equation*}
  \omega_-( \theta | \theta') = \begin{cases}
    u(\theta', 0) - u(\theta, +1), & \text{if\,  $u(\theta', 0) > u(\theta, +1)$;} \\
    0, & \text{otherwise},
  \end{cases}
\end{equation*}
and
\begin{equation*}
  \omega_+( \theta | \theta') = \begin{cases}
    u(\theta', +1) - u(\theta, 0), & \text{if\,  $u(\theta', +1) > u(\theta, 0)$}; \\
    0, & \text{otherwise}.
  \end{cases}
\end{equation*}
Using the utility function of our model, Eq.~(\ref{payoff-function}),
these expressions become
\begin{equation}
  \omega_-(\theta | \theta') = \begin{cases}
    - \theta' \, s(n_c) +\tilde{c},  & \text{if\,  $\theta' < c/s(n_c); $} \\
    0, & \text{otherwise},
  \end{cases} \label{wcondm}
\end{equation}
and
\begin{equation}
  \omega_+(\theta| \theta') = \begin{cases}
    -\tilde{c} + \theta\, s(n_c),  & \text{if\,  $\theta > c/s(n_c)$}; \\
    0, & \text{otherwise}.
  \end{cases}\label{wcondp}
\end{equation}
Now, inserting Eqs.~(\ref{wcondm}) and (\ref{wcondp}) into
Eqs.~(\ref{trans_totm}) and (\ref{trans_totp}) choosing $g(\theta)$ to be
a uniform distribution, we get
\begin{eqnarray}
  \omega_- ( \theta) &=& \epsilon\nonumber \\ &+ &
  \int_{\theta - \Delta \theta}^{\tilde{c}/s(n_c)} \left( - \theta' \, s(n_c) +\tilde{c}  \right) \, \frac{n(\theta',0)}{2 \Delta\theta}\, d\theta'  \label{zz1}\\
  \omega_+ (\theta) &=& \epsilon\nonumber \\ &+ &  \begin{cases}
    \left( -\tilde{c} + \theta\, s(n_c) \right)  \frac {n_c} {2 \Delta \theta} & \text{if\,  $\theta > \tilde{c}/s(n_c)$} \\
    0, & \text{otherwise} \end{cases}
  \label{zz2}
\end{eqnarray}
We emphasize that, in the presence of other distributions for the idiosyncratic term, the transition rates become more sophisticated, thus
indicating the non-physical nature of this dynamics.

\subsubsection{Logit dynamics}

When considering bounded rational agents, economics literature often
assumes that they do not imitate their neighbors, but follow a strategy
or action that would yield the best payoff for them. In line with this assumption,
one possible rule would be to change the action into cooperative
($\omega_+$) or defective ($\omega_-$) state with a transition rate
\begin{eqnarray}
  \omega_{\pm}  (  \theta ) &=& \frac{1}{1+\exp \left[ \mp \beta
\left( u(\theta,1) - u(\theta, 0 ) \right) \right]}.  \label{eq:logit} \end{eqnarray}
It is important to note that in this case the agent does not compare her
payoff with that of another agent, but with the payoff she would
obtain by using the opposite action. As there is no other agent involved,
there is also no interaction term in the above equation, which makes the
transition rates much simpler than in the previous case. This will be
advantageous for an analytical approach as we will see below.

The parameter $\beta$ in Eq.~(\ref{eq:logit}) quantifies the randomness
in the process: When $\beta$ is small, the agent is more likely to select
another action at random, even if that action is not more successful. On
the other hand, when $\beta\to\infty$, the rule becomes deterministic,
and the action that yields the maximum payoff is always chosen, as
posited by Ellison \cite{ellison:1993} when introducing his (myopic) best
response rule.

\section{Results}
\label{sec:3}

\subsection{Setup for computer simulations}
\label{setup}

In order to present our results in a clear manner, we will deal first
with the original model as introduced in \cite{spichtig:2011}, without
considering diversity nor external forcing. This will be the baseline
scenario against which we will subsequently illustrate the effects of
diversity to proceed to our main result, namely the influence of an
external driver and the concomitant appearance of diversity-induced resonance.

As described in the preceding Section, the model has several parameters
to specify. We start by measuring utilities as a function of the cost of
contributing to the public good, i.e., by taking $c=1$. For the
multiplication factor we fixed $r=5$ which, in a population of many
agents, is too small to induce agents to contribute to the public good.
Therefore, without the third term in Eq.~(\ref{payoff-function})
referring to the social norm, the only evolutionarily stable strategy
is defection. For the population size, we have chosen $N=10^3$ agents
(some runs were repeated with $N=10^4$ for the sake of comparison, yielding the same results).

Subsequently, we have chosen the following parameter values related to the social
norm. The strength of the norm is given by the slope $\alpha$ which, in
the absence of an external influence, is set as a constant $\alpha=1$,
albeit changes of this parameter do not qualitatively modify our
conclusions. Finally, for the sensitivity to the norm, we need to specify
the parameters of the distribution $g(\theta)$. In the following, we
consider two cases: (a) There is a sensitivity to the social norm equal
for all agents, which is given by the mean value $\Theta$ of the
distribution (homogeneous model). We will choose different values of
$\Theta$. (b) The sensitivity to the social norm is different for all
agents and randomly chosen from a uniform distribution in
$[\Theta-\sqrt{3} \Delta\theta, \Theta+\sqrt{3} \Delta\theta]$, where
$\Delta\theta$ is the standard deviation (heterogeneous model). Note
that our choice allows for negative sensitivities with effects as
described in Sect.~\ref{sec:2}.

To monitor the evolution of the system, we have measured the time-dependent density of
cooperators $n_{c}(t)= (1/N) \sum_i \sigma_i(t)$. To determine the stationary level of cooperation,
we compute the time-average number of cooperation, $n_{c} = \langle n_{c}(t)
\rangle_t$.  Subsequently, we also compute the  second moment of $n_c(t)$, i.e.~$\xi^2
= \langle \left( n_{c}(t) - n_{c} \right)^2 \rangle_t$, which is the
susceptibility of the system.

\subsection{Dynamics in the unforced model}\label{sec:unf}

\subsubsection{Model without diversity}

\begin{figure}[t]
\begin{center}
    \includegraphics[width=.48\textwidth]{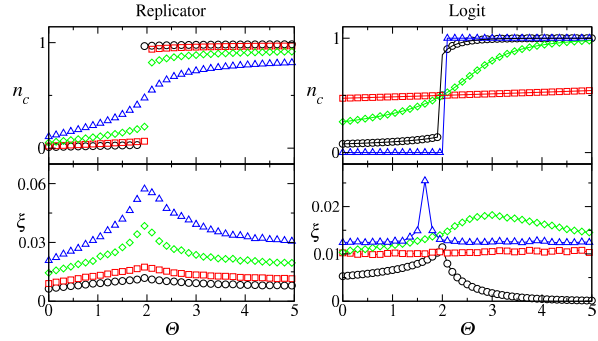}
    \caption{(upper row) Asymptotic fraction of cooperators $n_{c}$ dependent on the
      sensitivity to the social norm, $\theta \equiv\Theta$, which is equal for all agents in
      the model without diversity. (lower row) Fluctuations
      $\xi$ around the expected action, which is free-riding ($\sigma=0$)
      for $\Theta\ < 2$ and cooperation for $\Theta>2$.
(left column) Replicator dynamics.
      Different curves correspond to different values of $\epsilon$:
      triangles (0.01), diamonds (0.02), squares (0.05), circles (0.10).
(right column) Logit dynamics. Different curves correspond to
      different values of $\beta$: triangles (0.1), circles (1.0),
      diamonds (2.5), squares (10). The other parameters are described in the main text. In the
      upper right panel, the lines correspond to the analytical
      treatment, developed in section \ref{sec:nueva}.  }\label{figure1}
  \end{center}
\end{figure}

In the homogeneous model, the sensitivity to the social norm is equal for
all agents, $\theta\equiv \Theta$. Starting from an initial condition
where half of the population acts as cooperators, and half as
free-riders, Figure \ref{figure1} shows the asymptotic results of
computer simulations for the two update dynamics introduced in
Sect.~\ref{evol_dyn}. As it can be clearly seen in the upper panels, an increase
in the parameter $\Theta$ --that controls the influence of the social norm-- results in an
increase in the density of cooperators. For the replicator dynamics, and for
large values of randomness $\epsilon$, this effect becomes less visible as the width of the
transition increases. The results for the logit dynamics point in the
same direction, with $\beta^{-1}$ being the parameter that controls the
randomness or the frequency of mistakes.

The results of both dynamics become very similar when noise is very
small.  Note that Fig.~\ref{figure1} is obtained for equal initial
densities of contributors and free riders, but extensive simulations show that the
value of $\Theta$ at which the transition occurs does not depend on the
initial condition.
It is interesting to note the peak of the susceptibility (lower panels)
close to the transition towards cooperation, both for replicator and
logit dynamics. This is reminiscent of bistable systems which change their
stability at the transition. It is the archetypal situation where
diversity-induced resonance has already been demonstrated and, as we will
see below, it will give rise to the same behavior in this socio-economic
context.

\subsubsection{Model with diversity}

\begin{figure}
\begin{center}
\includegraphics[width=.48\textwidth]{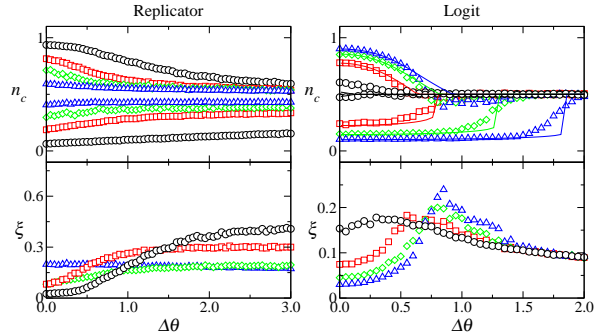}
\caption{(upper row) Asymptotic fraction of cooperators $n_{c}$ and
  (lower row) Susceptibility $\xi^{2}$ dependent on the sensitivity to
  the social norm, $\theta$, different for all agents in the model with
  diversity. $\Delta \theta$ is the variance of the distribution
  $g(\theta)$ with mean value $\Theta=2$.  (left column) Replicator
  dynamics.  Different curves correspond to different values of
  $\epsilon$: circles (0.02), squares (0.05), diamonds (0.07), triangles
  (0.10) (right column) Logit dynamics. Different curves correspond to
  different values of $\beta$: circles (2.0), squares (2.25), diamonds
  (2.50), triangles (2.75). In the upper panels, we have selected two different initial conditions ($n_c(0) = 0.1$ and $ 0.9$) for both kinds of dynamics.
  The other parameters are described in the main text. In the upper
  right panel, the lines correspond to the analytical treatment,
  developed in section \ref{sec:nueva}.}
\label{figure2}
\end{center}
\end{figure}

\begin{figure*}[ht]
\begin{center}
\includegraphics[width=.805\textwidth]{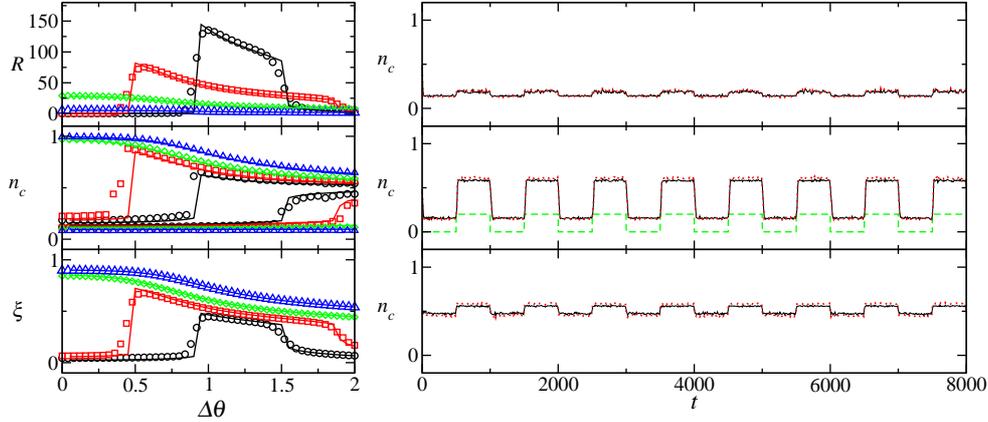}
\caption{Response of the system in presence of a periodic square-wave forcing with logit dynamics. In all the plots, $\beta=2.5$. Left column, first row: spectral amplification factor $R$, Eq.~(\ref{eq:saf}). Left column, middle row, maximum and minimum levels of cooperation
attained during
the evolution  of $n_c$  for the system. Left column, lowermost row,  the susceptibility.
Each symbol corresponds to a different signal amplitude: $\Delta\alpha = 0.05, 0.1, 0.2, 0.5$ (circle, square, diamond and triangle symbols respectively). Analytical results (see main text) are represented with solid lines.
In the right column, we depict the time dependency of the macroscopic state $n_c$
(solid, black lines), for three different values of the parameter $\Delta\theta$. The
values are $\Delta \theta = 0.7, 1.2, 1.7$ in the upper, middle and lower plot,
respectively. The dotted line represents the social pressure, while the thin
line (green on-line, only in the middle plot) shows the signal applied (not in the same scale, for
clarity).
Other parameters are: $T=10^3$, $N=10^4$, $r=5$, $\Theta = 2$, $\alpha = 1$.  } \label{response-myopic}
\end{center}
\end{figure*}

Using the results from the model without diversity as a reference case,
we now focus on the role of diversity in the sensitivity to the social
norm. That means that instead of a fixed value $\theta$ we consider an
individual value for each agent which is drawn from the uniform
distribution $g(\theta)$ specified in Sect.~\ref{setup}. The standard
deviation $\Delta\theta$ varies the degree of diversity.
The results of computer simulations are shown in Figure \ref{figure2}.
From the previous discussion (cf.~Fig.~\ref{figure1}) we know that, for the chosen set of parameters,
the transition from free riding to cooperation occurs at a value
$\Theta=2$. Therefore, in all the curves of Figure \ref{figure2}, we have fixed the average sensitivity to this value, in order to
investigate the role of diversity.
When plotting the stationary number of cooperators, the two curves for the same parameter set correspond to different initial conditions with a majority of cooperators or defectors.
From the simulation results, we can
clearly conclude that diversity alone does not favor the transition
towards cooperation. Also, increasing noise does not enhance this
situation.  From Fig.\ \ref{figure2} we see that, for the replicator
dynamics, the lower the noise the lower the cooperation in the asymptotic
state, reaching the random level of $n_c=0.5$ for very high values
(agents make mistakes every other time step on average).  For logit
dynamics the results are similar, but for low noise we observe an
asymmetric bifurcation in which the stationary state of low cooperation
merges onto the $n_c=0.5$ state only for very large values of the
diversity; higher noise values change the bifurcation toward a more
symmetric form.

\subsection{Dynamics under driving}

So far we have only discussed the role of the idiosyncratic sensitivity $\theta$ to the
social norm and found that it does not induce a transition to
cooperation. Now, as an important new ingredient, we consider that the
influence of the norm changes in time, expressed by the time-dependent
parameter $\alpha(t)$. Basically, any time dependence can be
considered. For simplicity we have chosen a periodic function in the form
of a square wave defined as
\begin{equation}
\alpha(t)=
\begin{cases}
  \alpha+\Delta\alpha,  & \text{if\,  $2nT<t<(2n+1)T$}; \\
  \alpha-\Delta\alpha, & \text{if\, $(2n+1)T<t<2(n+1)T$,}
\end{cases}\label{anxo1}
\end{equation}
with $n=0,1,2,\dots$.
In an adiabatic limit, where the period $T$ is large such that the system reaches the stationary equilibrium in a period, this situation is equivalent to the application of a single change in the social pressure as perceived by individuals (by external means, like a change of policy, for example).
We have verified that using a sinusoidal function
basically leads to the same results, qualitatively, than those shown in this Paper. So, we will focus on the expression given in Eq.~(\ref{anxo1}).

We already defined the global density of cooperators $n_{c}(t)$ to be
used as the order parameter. In particular, in the following, we will instead plot both the minimum and maximum values reached by the density of cooperators over time.  To further quantify the collective response
of the system to the externally changing influence of the social norm, we
introduce the Spectral Amplification Factor (SAF), $R$, defined as
\cite{gammaitoni:1998}
\begin{equation}\label{eq:saf}
  R=4 \frac{ |\langle n_{c}(t) {\rm e}^{i 2 \pi t/T}\rangle_t  |^2 }
{\Delta \alpha^2}.
\end{equation}
Now, in addition to the variance $\Delta \theta$ of the sensitivity to
the social norm, which describes an \emph{individual} feature, we further
have the change $\Delta \alpha$ in the social pressure caused by
\emph{external} influences.

\begin{figure}[t]
\begin{center}
\includegraphics[width=.35\textwidth]{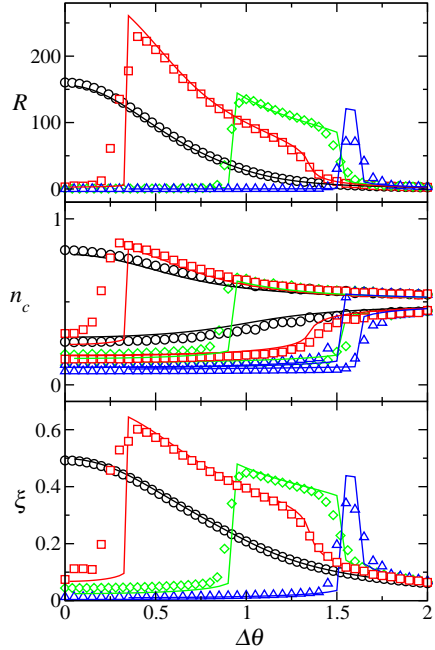}
\caption{Response of the system in presence of a periodic forcing with
  logit dynamics. In all the plots, $\Delta\alpha=0.05$. First row,
  spectral amplification factor $R$, Eq.~(\ref{eq:saf}). Middle row,
  maximum and minimum levels of cooperation attained during the evolution
  of $n_c$ for the system. Lowermost row, the system susceptibility.
  Each symbol (color) corresponds to a different value of the inverse
  randomness: $\beta = 2, 2.32, 2.5, 2.75$ (circles, squares, diamonds
  and triangles respectively).  With symbols we represent the results
  obtained by means of computer simulations, while the analytical results
  are presented with solid lines.  Other parameters are: $T=10^3$,
  $N=10^4$, $r=5$, $\Theta = 2$, $\alpha = 1$.
}\label{response-myopic-dalpha}
\end{center}
\end{figure}
\begin{figure}[t]
\begin{center}
\includegraphics[width=.35\textwidth]{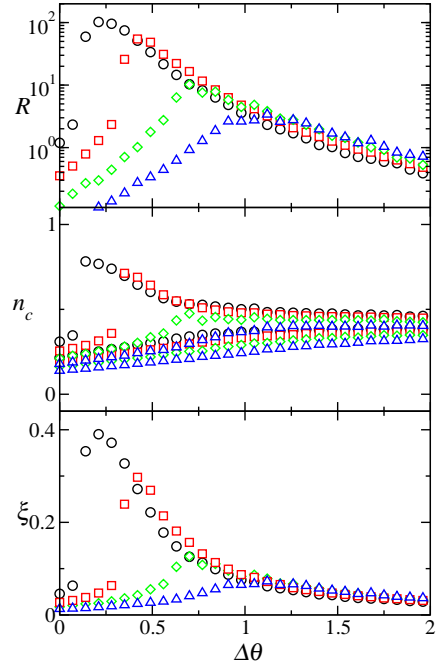}
\caption{Effect of a periodic signal applied to the social norm as a
  function of the population diversity $\Delta \theta$ for the replicator
  dynamics. In all the plots, $\Delta\alpha=0.05$. First row, spectral
  amplification factor $R$, Eq.~(\ref{eq:saf}). Middle row, maximum and
  minimum levels of cooperation attained during the evolution of $n_c$
  for the system. Lowermost row, the system's susceptibility.  Each
  curve corresponds to a different value randomness: $\epsilon = 0.04,
  0.045, 0.05, 0.055$ (circles, squares, diamonds and triangles
  respectively).  Other parameters are: $T=10^3$, $N=10^4$, $r=5$,
  $\theta = 2$, $\alpha = 1$.  }\label{response-repl}
\end{center}
\end{figure}

A summary of our numerical results is presented in Fig.\
\ref{response-myopic}. The left column  shows the spectral amplification factor $R$,
the maximum and minimum values of cooperation
$n_{c}$ and $\xi$ as a function
of the standard deviation of the diversity,
$\Delta\theta$, for different values of the amplitude of the external driving,
$\Delta\alpha$.  These results correspond solely to the logit dynamics; the
results for the replicator dynamics are qualitatively similar to those presented in the
plot and not shown.  We further noticed that, for all choices of
parameters, the results are independent of the initial conditions.
As it can be seen from the plots, for low $\Delta \theta$, the
response $R$ is largely independent of $\Delta \theta$. In this limit, it is possible to see the existence of super-threshold signal intensities $\Delta\alpha$, which are those values exhibiting large oscillations in the limit $\Delta\theta \to 0$. For the parameters in the plot, this corresponds to $\Delta\alpha \geq 0.2$.
On the other hand, for smaller values of signal amplitude, we find that (in the limit of small heterogeneity) the system responds simply linearly to changes in
the social norm. From a dynamical point of view, responses of the system to the external influence for low $\Delta\theta$ are depicted in the right column, upper panel.

However, intermediate values of $\Delta\theta$ do
provide evidence for resonant behavior if the driving intensity is
small: $R$ shows a peak for values of $\Delta\theta\simeq 1$,
which becomes more noticeable for smaller signals.
The oscillations of $n_c(t)$ are centered around $1/2$, a value much larger than the one obtained for lower values of diversity. Moreover,  the application of a one-time raising in the strength of the external signal may yield a non-linear response in terms of the growth of cooperating individuals.
From a policy making point of view, this translates in low incentive costs being able to enforce the cooperative state throughout the population.
When the driving amplitude $\Delta\alpha$ is
much larger, the system may be able to follow the signal simply because the signal is super-threshold, and the same thing happens even in absence of diversity.
Therefore, there is a true
resonance phenomenon, which can be observed for low external signals,
that elicits a strong response. In the middle panel of the right column, we show the dynamic response for a small applied signal, showing the large excursions in the number of cooperators when successively activating and de-activating the external signal.

Finally, for very large values of diversity, no response to the external influence is observed. The amount of individuals with very heterogeneous responses to the external signal does not allow a significant portion of the population to react to the external signal, and the system's response becomes linear again. The latter result can be observed in a vanishing response $R$, small oscillation amplitudes in $n_c$ (the latter, in the lower panel of the right column of plots).
In all the previous analyses, it is worth noticing that a peak in the susceptibility signals also the diversity-induced resonance in this system.

Analyzing the role of noise for a fixed driving strength $\Delta \alpha$,
we observe another interesting feature of the dynamics under driving.
Figs.\ \ref{response-myopic-dalpha} and \ref{response-repl} show for both
the replicator and the logit dynamics the appearance of stochastic resonance
\cite{gammaitoni:1998}. That means, for an intermediate noise intensity (temperature or randomness, in the proposed dynamics)
the diversity-induced resonance peak is more clearly observed, whereas
smaller or larger values of the randomness mostly suppress it. As with
most stochastic phenomena, the resonant behavior is clearly marked also
in the fluctuations of the system.
The observation of stochastic resonance is remarkable because it shows up
not in a physical, but in a socio-economic context. It indicates that some
level of imperfections in the adoption of the better performing
strategies may lead to larger responses to external stimuli.

It is also worth mentioning that the presence of contrarians, i.e.~of
agents which defect even in presence of social pressure, can be
beneficial for finding the diversity-induced resonance phenomenon. In some extreme cases, as shown in
Fig.~\ref{response-myopic-dalpha}, the resonance peak may appear only in
the presence of contrarians for $\beta=2.75$ \footnote{Assuming a uniform distribution of
  the sensitivity $\theta$ contrarians can be found in the population if
  $ \sqrt{3} \Delta\theta > \Theta $.}. This finding is against our
intuition that contrarians would hamper the adoption of a cooperative
state in the system. It reminds on the positive influence of destructive
agents on the emergence of cooperation in social dilemma situations as
discussed in \cite{Arenas2011113}, where this phenomenon was termed `the
joker effect'.

\section{Analytical approach}
\label{sec:nueva}

\subsection{Dynamics without driving}
\label{sec:nodrive}

To further understand the phenomenon of diversity-induced resonance in
our model, we now develop an analytical approach that should be compared
to the numerical simulations presented in the previous section.  While
the transition rates for the replicator dynamics, Eqs.~(\ref{zz1}) and
(\ref{zz2}), are too complicated for a tractable analytical approach, the
situation is different  for the logit dynamics.  In this case, the
density of agents with a given sensitivity $\theta$ depends only on the
total number of cooperators in the population, which is a macroscopic
variable. Consequently, with the transition rates of Eq.~(\ref{eq:logit})
and the equilibrium condition for the payoff function,
Eq.~(\ref{payoff-function}), we find for the transition rate towards the
cooperative (and defective) states the following expression:
\begin{eqnarray}
  \omega_{\pm}  (  \theta ) &=& \frac{1}{1+\exp
    \left\{ \mp \beta \left[ \tilde{c} - \theta \, s(n_c) \right]
    \right\}}.
  \label{eq:logit2}
\end{eqnarray}
From the above equation, we can trivially compute the density of cooperators
by integrating over the complete population of individuals,
\begin{eqnarray}
  n_c &=& \int d\theta' \, g(\theta') \frac{1}{1+\exp
    \left\{ \mp \beta \left[ \tilde{c} - \theta' \, s(n_c) \right] \right\}},
  \label{p:logit2}
\end{eqnarray}
which, by using the uniform distribution of the sensitivity $\theta$,
reduces to
\begin{equation}
  n_c = \frac{1}{2\Delta\theta}\int_{\Theta - \Delta\theta}^{\Theta +
    \Delta\theta} d\theta'  \frac{1}{1+\exp
\left\{ \mp \beta \left[ \tilde{c} - \theta \, s(n_c) \right] \right\}}.
  \label{p:logit3}
\end{equation}
Expanding this equation, one readily obtains for the density of cooperators
\begin{eqnarray}
n_c  &=& s(n_c) + \frac{ \ln  \left[ 1 +
\exp \left\{ -\beta [\tilde{c} - s(n_c)(\theta- \Delta\theta) ]
\right\} \right] } {2 s(n_c) \, \beta \, \Delta\theta } \nonumber \\
& & -\frac{ \ln \left[ 1 + \exp \left\{ -\beta [\tilde{c} +
      s(n_c)(\theta- \Delta\theta)
] \right\} \right] } {2 s(n_c) \, \beta \, \Delta\theta}.
\label{p:logit}
\end{eqnarray}
This equation can be solved self-consistently to obtain the stationary
value of $n_c$. The corresponding results are shown as solid lines in
Figs.~\ref{figure1} and \ref{figure2}, right columns. We find a very good agreement
between the numerical simulations and the prediction of our analytical
approach, thus further supporting the validity of our results.

In agreement with our discussion in Sect.~\ref{sec:unf},
the system exhibits a pitchfork bifurcation. When increasing the control
parameter $\Delta\theta$, the solution $n_c = 1/2$ changes its stability
from unstable to stable, when the two branches (one with a majority of
cooperators, the other with a majority of free-riders) collapse in the
center point. As observed in the simulations --and now confirmed by the analytical treatment--, the solutions are
asymmetrical with respect to the stable point, with the lowest branch
being less dependent on the value of the control parameter.

This is key to understand the mechanism behind the diversity-induced
resonance phenomenon in this socio-economic system: For intermediate
values of the diversity $\Delta \theta$,
small perturbations are sufficient to overcome the separatrix, i.e.~
the unstable solution $n_c=1/2$ that divides the attractor basins of the
two stable solutions. Thus, a signal which is usually too small to cause
transitions between those states, can be sufficient to trigger such a transition near the bifurcation
point. Farther from this critical point, a small signal only
causes linear response of the system, around a stable fixed point. This
fully confirms the discussion of the numerical results for the system
with driving in the previous section.

\subsection{Relaxational dynamics with driving}
\label{sec:otra}

After considering the dynamics without driving in the previous section,
we now turn to the dynamics with driving to better understand the
response of the system to the external change of the norm. We note that the change in
the  density of cooperators after one state has been selected for
update, is given by
\begin{equation}
 n_c(t+\delta t) = n_c(t) + \frac 1 N \left< \sigma_i(t+\delta t) - \sigma_i(t)| \left\{\sigma(t) \right\} \right>.
\end{equation}
where $\left< \cdot \right> $ represents the ensemble average, which is
conditional on $\left\{\sigma(t) \right\}$, i.e. all those states that
did not change. Going over to small $\delta t \equiv 1/N$, we arrive at the
continuous dynamics:
\begin{equation}
\frac{ d n_c(t)}{d t} \cong \left< \sigma_i(t+\delta t) | \left\{\sigma(t) \right\} \right> - n_c(t) .\nonumber
\end{equation}
The expected value for the selected state $\sigma_{i}$ after  update
can be expressed as
\begin{equation}
  \left< \sigma_i(t+\delta t) | \left\{\sigma(t) \right\} \right>
  = \hbox{Prob}[\sigma_i(t+\delta t) = 1] .
\end{equation}
Without loss of generality, the probability that $\sigma_{i}(t+\delta t)$
is $+1$, is given by $(1 - \hbox{Prob}[+1 \to 0]) + \hbox{Prob}[0 \to
+1]$, which for this system is given by
\begin{equation}
 \left< \sigma_i(t+\delta t) | \left\{\sigma(t) \right\} \right> =  \int d\theta' \, g(\theta') \left( 1 - \omega_-(\theta) + \omega_+(\theta) \right).
\end{equation}
Restricting ourselves again to the particular case of the uniform
distribution for $\theta$ and logit dynamics, we have
\begin{eqnarray}
 \frac{ d n_c(t)}{d t} &= &  f(n_c) =  \frac 1 2 -n_c(t) \\
 & & - \frac{ \ln \left[ \cosh \left( \beta \tilde{c} + \beta
s(n_c)(\Delta\theta-\theta) \right) \right] } { 4 \beta \, s(n_c)   \Delta\theta
} \nonumber \\
& & + \frac{ \ln \left[ \cosh \left( \beta \tilde{c} - \beta
s(n_c)(\Delta\theta+\theta) \right) \right] } { 4\beta \, s(n_c)   \Delta\theta
}. \nonumber
\end{eqnarray}

If the external signal given by $\alpha(t)$ is slow enough, we can
determine $R$ by assuming that $n_{c}(t)$ reaches its stationary state
fast compared to changes in $\alpha$. Then, $n_c(t) = n_c^*( \alpha(t)
)$.  For a squared signal, the spectral amplification factor is simply
given by
\begin{equation}
  R(n_c^*) = \frac{ \pi \left( n_c^*( \alpha+\Delta\alpha )
      - n_c^*( \alpha-\Delta\alpha ) \right)^2 }{\Delta\alpha^2  }.
\label{Rstar}
\end{equation}
For this forcing, the average number of cooperators reduces to $n_c^* = [ n_c^*(
\alpha+\Delta\alpha ) + n_c^*( \alpha-\Delta\alpha ) ] /2$. Then, the
susceptibility can be computed as
\begin{eqnarray*}
\xi^2& =&  \int_0^{T/2} dt \,
\left( n_c^*( \alpha+\Delta\alpha ) - n_{c} \right)^2 + \\
&  &\int_{T/2}^T dt \,  \left( n_c^*( \alpha-\Delta\alpha ) - n_{c} \right) ^2
\end{eqnarray*}
from which we get for the susceptibility
\begin{equation}
\xi^{2} = \left( n_c^*( \alpha+\Delta\alpha ) - n_c^*( \alpha-\Delta\alpha ) \right)^2.
\end{equation}

Figures \ref{response-myopic} and \ref{response-myopic-dalpha} present a
comparison between the analytical  and numerical results. As
with the previous comparisons, the match is very satisfactory.  While our
socio-economic model is quite different from a physics model, the dynamic
observations have similar underlying mechanisms as known in physical
systems with diversity-induced resonance, which makes it possible to apply a
standard analytical approach. For the replicator dynamics, we cannot
apply the same techniques to calculate the observables. But the fact that
we find in the simulations similarities between the logit dynamics, for
which we have analytical confirmation, and the replicator dynamics,
allows us to conjecture similarities in the underlying mechanisms.

\section{Discussion and conclusion}
\label{sec:4}

In this Paper, we have studied a socio-economic model of cooperation, to
understand the effect of social pressure on the contribution to a public
good \cite{spichtig:2011}. We tried to point out analogies with the
phenomenon of diversity-induced resonance in bistable physical systems
reported in Ref.~\cite{tessone:2006}. This was to show that
methodological input from Physics can be beneficial for social sciences,
in particular with respect to the vast knowledge about complex nonlinear
dynamical systems. By adopting an already existing model, we avoided to
impose a physics inspired toy model that may not have fitted the modeling
paradigms of social sciences.

Our analytical and numerical results demonstrate that our approach has
been largely successful. Indeed, we found strong evidence of
diversity-induced resonance, i.e., of the fact that the response of the
system to a weak external signal is stronger in a certain range of the
parameters governing the disorder in the system. Importantly, such strong
signals are sub-critical, meaning that these alone would not be able to
drive a homogeneous system, whereas diversity on its own would lead to an
undesired behavior (in our case, to a decrease in
cooperation). Furthermore, we have pursued another analogy to a physical
phenomenon, namely stochastic resonance \cite{gammaitoni:1998}. We found
evidence that there is an optimal range of noise or randomness to obtain
the response of the system to the external signal.

It is most interesting to interpret the above results in terms of the
original socio-economic model. In that context, diversity means different
sensitivity to the influence of the social pressure towards behaving in a
cooperative manner. If an external signal is emitted (e.g., changing laws
or incentives by the government) that leads to changes of the social
pressure, the population will follow these directions only if its
corresponding sensitivity to such pressure is diverse, but not too little
or too much. Homogeneous populations will simply ignore the new norms
whereas very heterogeneous populations will end up behaving in some kind
of ``average'' manner that does not follow the change.  This is in
agreement with the fact that strongly homogeneous groups, such as gangs
or sects, are very insensitive to external influences trying to bring
them to contribute to the general welfare. In an optimally diverse
population, on the contrary, we would see that the most sensitive people
would abide social pressure and start contributing to the common good,
thus leading to an increment of the social pressure that pushes other
agents, and so forth.

In this context, it is important to stress that the
phenomenon is robust against the kind of dynamics considered for the
transition towards cooperation. This is particularly meaningful as the
two cases studied in our paper, i.e. replicator and logit dynamics,
correspond to two completely different approaches to decision making from
the agent's viewpoint. While the former is based on a social, imitative,
component, the second describes a purely strategic behavior, even a
myopic one. Finally, we have observed that in some cases the required
degree of heterogeneity for the appearance of the resonance leads to the
existence of contrarian individuals in the population, which would
benefit from going against the norm. This resembles the case of
diversity-induced resonance arising from repulsive interactions and
related results in social dilemmas, as mentioned in Sect.~\ref{sec:2}.

It is also worth noticing that the phenomenon of diversity-induced
resonance only uses a weak signal to obtain the desired results. Strong
signals would drive the population irrespective of its degree of
diversity, but the external effort of the ``driver'' has to be much
larger. This may be important for policy-making decisions where costly
interventions in the society are not desirable because their benefit may
in the end be smaller than the incurred cost. Of course, the requirement
of diversity implies that these easily implemented policies may not be
possible for all groups or societies, which in itself is another hint to
policy makers about the need to estimate costs prior to specific
interventions. It goes without saying that applications of these ideas in
real life may need more complete models. For instance, one could think of
endogenously generated norm changes, involving a feedback between actions
and utility functions, or including the affective dimension of agents
by considering their emotional response \cite{schweitzer:2010}.
On the other hand, applying these ideas to organizations may require a careful consideration of hierarchical effects \cite{cronin:2012}. Such
improved models would lead to results that would be much more amenable to
comparison with actual social group dynamics or even with specifically
designed experiments, and thus contribute to our knowledge of the
mechanics of social improvement. Work along these lines is in progress.

\section*{Acknowledgments}

C.J.T.\  acknowledges
financial support from Swiss National Science Foundation through grant CR12I1\_125298 and SBF (Swiss Confederation) through research project
C09.0055.
A.S.\ was supported in part by grants MOSAICO and PRODIEVO (Ministerio de Ciencia e Innovaci\'on,
Spain), RESINEE from ERA-Net on Complexity,
and MODELICO-CM (Comunidad de Madrid, Spain). A.\ S.\ is thankful to the
Chair of Systems Design at ETH Z\"urich for the warm hospitality
enjoyed during the design of this work.


\begin{thebibliography}{10}

\bibitem{castellano:2009}
Claudio Castellano, Santo Fortunato, and Vittorio Loreto.
\newblock Statistical physics of social dynamics.
\newblock {\em Rev. Mod. Phys.}, 81(2):591--646, 2009.

\bibitem{stauffer:2003}
D.~Stauffer.
\newblock Sociophysics simulations.
\newblock {\em Computing in Science Engineering}, 5(3):71 -- 75, 2003.

\bibitem{tessone:2006}
Claudio~J. Tessone, Claudio~R. Mirasso, Ra\'ul Toral, and James~D. Gunton.
\newblock Diversity-induced resonance.
\newblock {\em Phys. Rev. Lett.}, 97(19):194101, 2006.

\bibitem{tessone:2007}
C.~J. Tessone, A.~Scir\`e, R.~Toral, and P.~Colet.
\newblock Theory of collective firing induced by noise or diversity in
  excitable media.
\newblock {\em Phys. Rev. E}, 75(1):016203, 2007.

\bibitem{toral:2007}
R.~Toral, C.~J. Tessone, and J.~V. Lopes.
\newblock Collective effects induced by diversity in extended systems.
\newblock {\em Eur. Phys. J. Special Topics}, 143:59--67, 2007.

\bibitem{Chen2009}
Hanshuang Chen, Yu~Shen, Zhonghuai Hou, and Houwen Xin.
\newblock {Resonant response of forced complex networks: the role of
  topological disorder.}
\newblock {\em Chaos (Woodbury, N.Y.)}, 19(3):033122, 2009.

\bibitem{Chen2009a}
Hanshuang Chen, Zhonghuai Hou, and Houwen Xin.
\newblock {Threshold-diversity-induced resonance}.
\newblock {\em Physica A: Statistical Mechanics and its Applications},
  388(12):2299--2305, 2009.

\bibitem{Komin2010}
Niko Komin, Lucas Lacasa, and Ra\'{u}l Toral.
\newblock {Critical behavior of a Ginzburg--Landau model with additive quenched
  noise}.
\newblock {\em Journal of Statistical Mechanics: Theory and Experiment},
  2010(12):P12008, 2010.

\bibitem{Calisto2010}
H~Calisto and M~G Clerc.
\newblock {A new perspective on stochastic resonance in monostable systems}.
\newblock {\em New Journal of Physics}, 12(11):113027, 2010.

\bibitem{Perez2010}
Toni P\'{e}rez, Claudio~R Mirasso, Ra\'{u}l Toral, and James~D Gunton.
\newblock {The constructive role of diversity in the global response of coupled
  neuron systems.}
\newblock {\em Philosophical transactions. Series A, Mathematical, physical,
  and engineering sciences}, 368(1933):5619--32, 2010.

\bibitem{WANG2010}
Qing~Yun Wang, Matja\v{z} Perc, Zhi~Sheng Duan, and Guan~Rong Chen.
\newblock {Spatial Coherence Resonance In Delayed Hodgkin--huxley Neuronal
  Networks}.
\newblock {\em International Journal of Modern Physics B}, 24(09):1201, 2010.

\bibitem{Wu2011}
Dan Wu, Shiqun Zhu, and Xiaoqin Luo.
\newblock {Diversity-induced resonance with two different kinds of time
  delays}.
\newblock {\em Physica A: Statistical Mechanics and its Applications},
  390(11):1835--1840, 2011.

\bibitem{McDonnell2011}
Mark~D McDonnell and Lawrence~M Ward.
\newblock {The benefits of noise in neural systems: bridging theory and
  experiment.}
\newblock {\em Nature reviews. Neuroscience}, 12(7):415--26, 2011.

\bibitem{tessone:2009}
Claudio~J. Tessone and Ra\'ul Toral.
\newblock Diversity-induced resonance in a model for opinion formation.
\newblock {\em Eur. Phys. J. B}, 71(4):549 -- 555, 2009.

\bibitem{galam:1997}
Serge Galam.
\newblock Rational group decision making: A random field ising model at t = 0.
\newblock {\em Physica A}, 238(1-4):66 -- 80, 1997.

\bibitem{sethna:1993}
James~P. Sethna, Karin Dahmen, Sivan Kartha, James~A. Krumhansl, Bruce~W.
  Roberts, and Joel~D. Shore.
\newblock Hysteresis and hierarchies: Dynamics of disorder-driven first-order
  phase transformations.
\newblock {\em Phys. Rev. Lett.}, 70(21):3347--3350, 1993.

\bibitem{VazMartins2010}
T.~{Vaz Martins}, M.~Pineda, and R.~Toral.
\newblock {Mass media and repulsive interactions in continuous-opinion
  dynamics}.
\newblock {\em EPL (Europhysics Letters)}, 91(4):48003, 2010.

\bibitem{Deffuant2000}
Guillaume Deffuant, David Neau, Frederic Amblard, and G\'{e}rard Weisbuch.
\newblock {Mixing beliefs among interacting agents}.
\newblock {\em Advances in Complex Systems}, 3:87--98, 2000.

\bibitem{Tessone2008}
Claudio~Juan Tessone, Dami\'{a}n~H Zanette, and Ra\'{u}l Toral.
\newblock {Global firing induced by network disorder in ensembles of active
  rotators}.
\newblock {\em The European Physical Journal B}, 62(3):319--326, 2008.

\bibitem{Tessone2012}
{Tessone, Claudio J.} and {Zanette, Dami\'an H.}
\newblock Synchronised firing induced by network dynamics in excitable systems.
\newblock {\em Europhysics Letters}, 2012.
\newblock To appear.

\bibitem{PhysRevE.81.041103}
T.~Vaz~Martins, V.~N. Livina, A.~P. Majtey, and R.~Toral.
\newblock Resonance induced by repulsive interactions in a model of globally
  coupled bistable systems.
\newblock {\em Phys. Rev. E}, 81:041103, 2010.

\bibitem{PhysRevE.85.011150}
Georges Harras, Claudio~J. Tessone, and Didier Sornette.
\newblock Noise-induced volatility of collective dynamics.
\newblock {\em Phys. Rev. E}, 85:011150, 2012.

\bibitem{GALAM2004}
S~Galam.
\newblock {Contrarian deterministic effects on opinion dynamics: ``the hung
  elections scenario''}.
\newblock {\em Physica A: Statistical and Theoretical Physics}, 333:453--460,
  2004.

\bibitem{Stauffer2004558}
D.~Stauffer and J.S.~S{\'a} Martins.
\newblock Simulation of galam's contrarian opinions on percolative lattices.
\newblock {\em Physica A: Statistical Mechanics and its Applications},
  334(3-4):558 -- 565, 2004.

\bibitem{Arenas2011113}
Alex Arenas, Juan Camacho, Jos{\'e}~A. Cuesta, and Rub{\'e}n~J. Requejo.
\newblock The joker effect: Cooperation driven by destructive agents.
\newblock {\em Journal of Theoretical Biology}, 279(1):113 -- 119, 2011.

\bibitem{spichtig:2011}
Mathias Spichtig and Christian Traxler.
\newblock Social norms and the indirect evolution of conditional cooperation.
\newblock {\em Journal of Economics}, 102:237--262, 2011.

\bibitem{keser:2000}
Claudia C.~Keser and F.~van Winden.
\newblock Conditional cooperation and voluntary contributions to public goods.
\newblock {\em Scand. J. Econ.}, 102:23, 2000.

\bibitem{fischbacher:2001}
S.~G\"achter U.~Fischbacher and E.~Fehr.
\newblock Are people conditionally cooperative? evidence from a public goods
  experiment.
\newblock {\em Econ. Lett.}, 71:397, 2001.

\bibitem{rapoport:1966}
Anatol Rapoport and Melvin Guyer.
\newblock A taxonomy of $2 \times 2$ games.
\newblock {\em General Systems}, 11:203--214, 1966.

\bibitem{axelrod:1984}
R.\ Axelrod.
\newblock {\em The Evolution of Cooperation}.
\newblock Basic Books, New York, 1984.

\bibitem{kagel:1995}
John~H. Kagel and Alvin~E. Roth.
\newblock {\em The Handbook of Experimental Economics}.
\newblock Princeton University Press, Cambridge, Massachusetts, 1995.

\bibitem{gachter:2007}
S.~G\"achter.
\newblock Conditional cooperation: Behavioral regularities from the lab and the
  field and their policy implications.
\newblock In B.~S. Frey and A.~Stutzer, editors, {\em Economics and Psychology.
  A Promising New Cross-Disciplinary Field}. MIT Press, 2007.

\bibitem{grujic:2010}
Jelena Gruji\'c, Constanza Fosco, Lourdes Araujo, Jos\'e~A. Cuesta, and Angel
  S\'anchez.
\newblock Social experiments in the mesoscale: Humans playing a spatial
  prisoner's dilemma.
\newblock {\em PLoS ONE}, 5(11):e13749, 2010.

\bibitem{grujic:2012}
Jelena Gruji{\'c}, Burcu Eke, Antonio Cabrales, Jos{\'e}~A. Cuesta, and Angel
  S{\'a}nchez.
\newblock {\em Sci. Rep.}, 2:638, 2012.

\bibitem{gracia-lazaro:2012}
Carlos Gracia-L{\'a}zaro, Alfredo Ferrer, Gonzalo Ruiz, Alfonso Taranc{\'o}n,
  Jos{\'e}~A. Cuesta, Angel S{\'a}nchez, and Yamir Moreno.
\newblock {\em Proc. Natl. Acad. Sci. USA}, 109:12922--12926, 2012.

\bibitem{fehr:2006}
E.~Fehr and K.~Schmidt.
\newblock The economics of fairness, reciprocity and altruism experimental
  evidence and new theories.
\newblock In S.~C. Kolm and J.~M. Ythier, editors, {\em Handbook on the
  Economics of Giving, Reciprocity and Altruism, Vol.1}. North-Holland, 2006.

\bibitem{coleman:1990}
James Coleman.
\newblock {\em Foundations of social theory}.
\newblock Harvard University Press, Cambridge, Massachusetts, 1990.

\bibitem{galan:2005}
Jos\'{e}~Manuel Gal\'{a}n and Luis~R. Izquierdo.
\newblock Appearances can be deceiving: Lessons learned re-implementing
  axelrod's 'evolutionary approach to norms'.
\newblock {\em J. Artif. Societies Soc. Simulation}, 8(3):2, 2005.

\bibitem{mengel:2008}
F.~Mengel.
\newblock Matching structure and the cultural transmission of social norms.
\newblock {\em J. Econ. Behav. Org.}, 67:608, 2008.

\bibitem{ostrom:2000}
Elinor Ostrom.
\newblock Collective action and the evolution of social norms.
\newblock {\em J. Econ. Perspec.}, 14:137, 2000.

\bibitem{posner:2000}
E.~A. Posner.
\newblock {\em Law and Social Norms}.
\newblock Harvard University Press, 2000.

\bibitem{szabo:2007}
Gy{\"o}gy Szab{\'o} and G{\'a}bor F{\'a}th.
\newblock Evolutionary games on graphs.
\newblock {\em Phys.\ Rep.}, 446:97--216, 2007.

\bibitem{roca:2009a}
Carlos~P. Roca, Jos{\'e} Cuesta, and Angel S{\'a}nchez.
\newblock Evolutionary game theory: temporal and spatial effects beyond
  replicator dynamics.
\newblock {\em Phys. Life Rev.}, 6:208--249, 2009.

\bibitem{helbing:1992}
Dirk Helbing.
\newblock Interrelations between stochastic equations for systems with pair
  interactions.
\newblock {\em Physica A}, 181:29--52, 1992.

\bibitem{schlag:1998}
Karl~H.\ Schlag.
\newblock Why imitate, and if so, how? \mbox{A} boundedly rational approach to
  multi-armed bandits.
\newblock {\em J.\ Econ.\ Theory}, 78:130--156, 1998.

\bibitem{hofbauer:1998}
J.\ Hofbauer and K.\ Sigmund.
\newblock {\em Evolutionary Games and Population Dynamics}.
\newblock Cambridge University Press, Cambridge, 1998.

\bibitem{gintis:2009}
Herbert Gintis.
\newblock {\em Game theory evolving}.
\newblock Princeton University Press, Princeton, 2nd edition, 2009.

\bibitem{alos-ferrer:2010}
Carlos Al{\'o}s-Ferrer and Nick Netzer.
\newblock The logit-response dynamics.
\newblock {\em Games Econ. Behav.Economic Behavior}, 68:413--427, 2010.

\bibitem{ellison:1993}
Glenn Ellison.
\newblock Learning, local interaction, and coordination.
\newblock {\em Econometrica}, 61:1047--1071, 1993.

\bibitem{gammaitoni:1998}
Luca Gammaitoni, Peter H\"anggi, Peter Jung, and Fabio Marchesoni.
\newblock Stochastic resonance.
\newblock {\em Rev. Mod. Phys.}, 70(1):223--287, 1998.

\bibitem{schweitzer:2010}
{Frank Schweitzer} and {David Garc\'{\i}a}.
\newblock An agent-based model of collective emotions in online communities.
\newblock {\em Eur. Phys. J. B}, 77:533--545, 2010.

\bibitem{cronin:2012}
Katherine Cronin and Angel S{\'a}nchez.
\newblock {\em Adv. Complex Sys.}, 15 Suppl. no. 1:1250066, 2012.

\end{thebibliography}
\end{document}